\documentstyle[12pt]{article}

\begin{document}
\vbox to 18pt{}
\begin{center}
{\Large Diffusion in momentum space as a picture
of second-order Fermi acceleration}

\vspace{10mm}

{\large M. Ostrowski$^{1,2}$ \&  G. Siemieniec-Ozi\c{e}b{\l}o$^1$}
\end{center}

\noindent
{\small\em $^1$Obserwatorium Astronomiczne,
Uniwersytet Jagiello\'nski, ul.Orla 171, 30-244 Krak\'ow, \\ 
\indent Poland (Internet: mio@oa.uj.edu.pl \& grazyna@oa.uj.edu.pl) \\
$^2$Max-Planck-Institut f\"ur Radioastronomie, 
Postfach 2024, 53010 Bonn, Germany}

\vspace{3mm}

\begin{abstract}
Energetic particles in a turbulent medium can be subject to
second-order Fermi acceleration due to scattering on moving plasma
waves. This mechanism leads to growing particle momentum dispersion and,
at the same time, increases the mean particle energy. In the most
frequently met situations both processes can be represented by a
single momentum diffusion term in the particle kinetic equation. In the
present paper we discuss the conditions allowing the additional term for
regular acceleration to arise. For forward-backward asymmetric
scattering centres, besides the diffusive term one should explicitly
consider the regular acceleration term in momentum space, which can consist 
of the first-order ($\propto V$), as well as the second-order ($\propto
V^2$) part in the wave velocity $V$. We derive the condition for the
scattering probability in the wave rest frame required for vanishing the
regular acceleration term and provide a simple mechanical example
illustrating the theoretical concepts. Finally, we address its possible
role in cosmic ray acceleration processes. 
\end{abstract}

{\baselineskip 1.0\baselineskip
\section{Introduction}

The original theory conceived by Fermi [5] for the acceleration of
cosmic rays considers charged particles scattered by magnetic clouds
moving with speeds $V$ relative to the plasma rest frame. The resultant
acceleration is represented by a systematic gain term. The following
studies considered some particular acceleration scenarios involving
large amplitude magnetohydrodynamic waves and allowing for regular
acceleration of cosmic ray particles (cf. Parker [7]). A general
approach is based on the Fokker-Planck equation (cf. Parker \& Tidman
[8]; Clemmow \& Dougherty [2]; Lacombe [6]; Blandford \& Eichler [1]).
In the recent years this approach supplemented with the quasi-linear
derivation of transport coefficients for cosmic rays interacting with
Alfv\'enic turbulence (e.g. Schlickeiser [9]) became an analytical
basis for considering energetic particle transport phenomena in
rarefied astrophysical plasmas. Within this approach the detailed
balance principle applied to particle interaction with the field of
magnetized plasma waves leads to the reduction of the general
Fokker-Planck equation to the pure momentum diffusion equation. Let us
note, however, that the principle of detailed balance which arises in a
natural way in the kinetic description of gas of scattering particles
is no longer a natural assumption for particles scattered on external
`heavy' scattering centres. In our opinion this till now neglected fact
deserves more detailed study because of its potential importance for
cosmic ray particle acceleration in turbulent media.

In the present work we consider detailed conditions for the angular
scattering probability that is required for reduction of the
Fokker-Planck equation describing particle momentum scattering into the
momentum diffusion equation. The next section presents general
derivation leading to such simplification. Then, in section 3, we
present calculation of the Fokker-Planck coefficients for the
scattering centres (below we call them simply 'waves') moving with
velocities much smaller than the particle velocity. We show that the
scattering probability distribution, which is - in the scatterig wave
rest frame - symmetric with respect to the interchange of particle
momentum before and after scattering, is required to describe the
process as a pure momentum diffusion. A simple mechanical model
allowing the breaking this symmetry and yielding a regular acceleration
(deceleration) term is discussed in section 4. In the discussion
(section 5) we stress the importance of the possibility to have such
regular term in the stochastic acceleration process for cosmic ray
production in astrophysical sources.

\section{The Fokker-Planck approach}

As above mentioned the approach based on the Fokker-Planck equation is
generally used to describe energetic particle interactions with
magnetohydrodynamic turbulence. Following Blandford \& Eichler [1] we
write --- in the phase space --- the equation describing the evolution
of the isotropic part of particle distribution function $f({\bf r},
{\bf p}, t)$, due to momentum scattering as

$${\partial f \over \partial t} = {\bf \nabla}_p \cdot \left\{ - {\left<
\Delta {\bf p}\right> \over \Delta t} f  + {1 \over 2}\,{\bf \nabla}_p
\cdot \left[ {\left< \Delta {\bf p}\, \Delta {\bf p}\right> \over \Delta
t} f \right] \right\} \qquad , \eqno(2.1)$$

\noindent
where the Fokker-Planck coefficients, the mean momentum gain rate
${\left< \Delta {\bf p}\right> \over \Delta t}$ represents the Fermi
term and the rate for growing momentum dispersion is governed by the
term ${\left< \Delta {\bf p} \,\Delta {\bf p} \right> \over \Delta t}$;
${\bf \nabla}_p \equiv \partial / \partial {\bf p}$. With a probability
for changing a particle momentum in an individual scattering from ${\bf
p}$ to ${\bf p}^\prime$ ($\equiv {\bf p}+\Delta {\bf p}$) given by the
function $\Psi ({\bf p}^\prime, {\bf p})$ , with the normalization
$\int \Psi d\Delta {\bf p} = 1$, these coefficients are given as

$${\left< \Delta {\bf p} \right> \over \Delta t} = {1 \over \Delta t}
\int \Psi({\bf p}^\prime, {\bf p})({\bf p}^\prime-{\bf p}) d\Delta {\bf
p}  \eqno(2.2) $$

\noindent and

$${\left< \Delta {\bf p} \,\Delta {\bf p} \right> \over \Delta t} = {1
\over \Delta t} \int \Psi({\bf p}^\prime, {\bf p})({\bf p}^\prime-{\bf
p})^2 d\Delta {\bf p} \qquad , \eqno(2.3)$$

\noindent
where $1/\Delta t$ is the scattering frequency. If the
principle of detailed balance holds, $\Psi({\bf p}^\prime,{\bf p}) =
\Psi({\bf p}, {\bf p}^\prime)$, we have (Blandford \& Eichler [1]):

$${{\bf \nabla}_p} \left\{ {\left< \Delta {\bf p} \right> \over \Delta
t} - {1\over 2} \,{\bf \nabla}_p \cdot {\left< \Delta {\bf p} \,\Delta
{\bf p}\right> \over \Delta t} \right\}  = 0 \qquad . \eqno(2.4)$$

\noindent
The equilibrium momentum distribution can exist on the condition of
the momentum-change velocity vanishing (i.e. a constant vector in
parentheses in Eq.~2.4). Then, the Fokker-Planck coefficients
relate as

$${\left< \Delta {\bf p} \right> \over \Delta t} = {1\over 2} \,{\bf
\nabla}_p \cdot {\left< \Delta {\bf p} \,\Delta {\bf p} \right> \over
\Delta t} \qquad , \eqno(2.5)$$

\noindent
and Eq.~(2.1) takes a simple form

$${\partial f \over \partial t} = {\bf \nabla}_p \cdot \left\{ {\bf
D}_{pp} \cdot {\bf \nabla}_p f \right\} \qquad , \eqno(2.6)$$

\noindent
where one introduced the momentum diffusion tensor ${\bf D}_{pp} \equiv
\left< \Delta {\bf p}\, \Delta {\bf p}  \right> / ( 2 \Delta t ) $.
For the isotropic momentum diffusion ${\bf D}_{pp} = D_p\,{\bf I}$,
where $D_p$ is the momentum diffusion coefficient and ${\bf I}$ is the
diagonal unit matrix. Then Eq.~(2.1) transforms into the form with the
scalar momentum and the operator $\partial / \partial p$ replacing the
vector $\bf p$ and $\nabla_p$, respectively, and, accordingly,
Eq.~(2.6) takes the form

$${\partial f \over \partial t} = {1 \over p^2} {\partial \over \partial
p} \left\{ p^2 \, D_p \, {\partial f \over \partial p}\right\} \qquad .
\eqno(2.7)$$

\noindent
In the present work we consider detailed conditions required  
for reduction of Eq.~(2.1) into (2.7). 

\section{Fokker-Planck coefficients and the condition for vanishing
a regular acceleration term}

Any wave damping or boosting process in rarefied astrophysical plasmas
can be usually neglected for MHD waves at time scales relevant for
single act of energetic particle interaction with a given wave. Due to
this fact the electric field can be assumed to vanish in the wave rest
frame (`wave frame') and the transport coefficients in the momentum
space can be obtained using a straightforward procedure.  Let us
consider an individual wave (scattering centre) taken from the ensemble
of waves with isotropic distribution of velocities. In the wave frame
the scattering conserves the particle energy and can be described with
the angular scattering cross-section. Thus, the energy gain of any
individual particle can be derived from two Lorentz transformations,
the transformation from the mean plasma rest frame (`plasma frame') to
the wave frame before the scattering, and in the opposite direction
after the scattering. To derive transport coefficients one simply has
to calculate the change in momentum on scattering and to average it
over incoming and outgoing pitch angles. This approach was proposed by
Duffy [4] and the discussion in this section follows his work in
essentials.

Let us take the $z$-axis of the considered reference frame along the
mean background magnetic field. In the plasma frame we denote an $X$
quantity before interaction and after interaction with a prime,
$X^\prime$. In the wave frame an index $w$ is added to respective
quantities. Then, we denote with $V$ a wave velocity assumed to be
along the magnetic field\footnote{But the derivations are performed
with all angles measured with respect to the direction of the
scattering centre velocity and, in principle, has nothing to do with
the magnetic field.}, with $v$ being a particle velocity ($v \gg V$),
and with $p$ and $\mu$ a particle momentum and a pitch angle cosine,
respectively (below, $\mu$ is called shortly `a pitch angle'). The
Lorentz transformation of particle momentum into the wave frame gives,
to second order in $V/v$ (as required by the Fokker--Planck approach),

$$p_w = p \left[ 1 - {\mu V \over v} + {1 \over 2} \left(1-\mu^2 \right)
{V^2 \over v^2} + {\mu^2 \over 2} {V^2 \over c^2} \right] \eqno(3.1)$$

\noindent and, to the first order in $V/v$,

$$\mu_w = \mu - \left( 1-\mu^2 \right) {V \over v} \qquad . \eqno(3.2)$$

\noindent
The particle momentum change $\Delta p \equiv p^\prime - p$ at an
individual scattering transforming $\mu$ into $\mu^\prime$ is

$$\Delta p = \left[ \left( \mu^\prime-\mu \right) {V\over v} + \left(
{\mu^\prime}^2-\mu^2 \right) {V^2\over 2\,v^2} + \left( \mu^\prime-\mu
\right)^2 {V^2\over 2\,c^2} \right] \, p \qquad \eqno(3.3)$$

\noindent
where the terms of the higher-than-second order in $V/v$ are neglected.
Let us note that in the expression (3.3), there are no zero-order terms
in wave velocity. Therefore, in order to derive $<\Delta p>$ and
$<(\Delta p)^2 >$  valid up to the second-order in $V/v$ one should
derive the required weighting factors up to the first-order and to the
zero-order, respectively. The factor which bears all physical
information about the scattering is a probability for the particle pitch
angle $\mu$ to be transformed into $\mu^\prime$. The scattering process
is described here in the wave frame, where we consider the probability
$P_w(p_w,\mu_w,\mu_w^\prime)$ for a particle with momentum $p_w$ to
change $\mu_w$ to $\mu^\prime_w$ during a time $\Delta t_w$ (= $\Delta
t$ + terms $O(V^2/v^2)$ ). At a scattering the particle momentum $p_w$
is assumed to be conserved. To the first order the scattering frequency
in the plasma frame is proportional to the Lorentz factor

$${\gamma_w v_w \over \gamma \, v} = 1 - \mu {V \over v} \qquad ,
\eqno(3.6)$$

\noindent
where $\gamma \equiv 1/\sqrt{1-v^2}$. Thus, expressions for the
Fokker-Planck coefficients valid to the second order in $V/v$ are

$$ \left< \Delta p \right>  = {1 \over 4 }\, \int^1_{-1} d\mu
\int^1_{-1} d\mu^\prime\, \left( 1 - \mu {V \over v} \right) 
P(p,\mu, \mu^\prime)\,
\Delta p (p, \mu, \mu^\prime) \quad \eqno(3.8) $$

\noindent and

$$\left< (\Delta p)^2 \right> = {1 \over 4 }\, \int^1_{-1} d\mu
\int^1_{-1} d\mu^\prime\, P(p,\mu, \mu^\prime)\, \Delta p^2 (p, \mu,
\mu^\prime) \qquad , \eqno(3.9) $$

\noindent
where the probability $P$ should be expressed with the use of $P_w$, by
expanding it to the first-order in $V/v$. Let us note that the above
coefficients are the ones for the isotropic part of the distribution
function, denoted above as $f(p)$. In the wave frame, a number of
particle transitions ($\mu_w \rightarrow \mu_w^\prime$) per unit time is
proportional to $P_w(p_w, \mu_w, \mu_w^\prime)$ and to the slightly
anisotropic particle density $f_w(p_w,\mu_w) = f(p) \cdot ({\rm d}\mu /
{\rm d}\mu_w)$. Therefore, one can write the probability function in the
plasma frame as
         
$$P(p,\mu,\mu^\prime) = P_w \left[ p_w(p,\mu), \mu_w(\mu),
\mu_w^\prime(\mu^\prime) \right] {{\rm d}\mu_w^\prime \over {\rm
d}\mu^\prime_{ }} \qquad . \eqno(3.10)$$

\noindent
With the use of Eq-s (3.1,2)  the integrand in (3.8) can be expanded
to the first-order as

$$\left( 1 - \mu {V \over v} \right)\, P(p,\mu,\mu^\prime) =
P_w(p,\mu,\mu^\prime) \left( 1-\mu {V \over v} + 2 \mu^\prime {V \over
v} \right) - p \mu {V \over v} {\partial P_w(p,\mu,\mu^\prime) \over
\partial p} $$

$$- \left(1-\mu^2 \right) {V \over v} {\partial P_w(p,\mu,\mu^\prime)
\over \partial \mu} - \left(1-{\mu^\prime}^2\right) {V \over v}
{\partial P_w(p,\mu,\mu^\prime) \over \partial \mu^\prime} \qquad .
\eqno(3.11)$$

\noindent
In the integral in (3.9) we need analogous expansion to the zeroth-order
in $V/v$, $P(p, \mu, \mu^\prime) = P_w(p, \mu, \mu^\prime)$, to obtain
the momentum diffusion coefficient as

$$D_p = {1 \over \Delta t} 
{p^2 \over 8} \left( V \over v \right)^2 \int_{-1}^1 {\rm d} \mu
\int_{-1}^1 {\rm d} \mu^\prime \left( \mu^\prime - \mu \right)^2
P_w(p,\mu,\mu^\prime) \qquad . \eqno(3.12)$$

\noindent
In the above formula one can consider the transition probability in the
wave rest frame, where the cosmic ray scattering can be described in the
most simple way. For example, in the case of scattering on
magnetohydrodynamic waves the electric field vanishes in such a frame
leading to interactions conserving particle energies.

Let us consider the condition (2.5) required to reduce the acceleration
process to the purely diffusive one (Eq.~2.6,8). With the scattering
probability given in Eq.~(3.11), one can derive the mean momentum
change in Eq.~(3.8). For a general form of $P_w$ the regular
acceleration term should be appended to Eq.~(2.7) besides the momentum
diffusion term. Then, it takes a more general form

$${\partial f \over \partial t} = {1 \over p^2} {\partial \over \partial
p} \left\{ p^2 \, D_{pp} \, {\partial f \over \partial p}\right\} -
{1 \over p^2} {\partial \over \partial p} \left\{ p^2 \, \dot{p}_{reg} f
\right\} \qquad , \eqno(3.13)$$

\noindent
where $\dot{p}_{reg} \equiv \, <\Delta p>_{reg} / \Delta t$ is a regular
momentum changing term, equivalent to the regular acceleration if 
$\dot{p}_{reg} > 0$. With Eq-s~(2.5,3.8-9) one obtains

$${<\Delta p>_{reg} \over p} = -{1 \over 4} \left( V \over v \right)
\int_{-1}^1 {\rm d}\mu \int_{-1}^1 {\rm d}\mu^\prime \left( 
\mu^\prime - \mu \right) P_w(p,\mu,\mu^\prime) $$
$$ -{1 \over 4} \left( V \over v \right)^2 \int_{-1}^1 {\rm d}\mu
\int_{-1}^1 {\rm d}\mu^\prime \left( {\mu^\prime}^2 - \mu^2 \right)
P_w(p,\mu,\mu^\prime) $$
$$ +{1 \over 4} \left( V \over v \right)^2 \int_{-1}^1 {\rm d}\mu
\int_{-1}^1 {\rm d}\mu^\prime \left( \mu^\prime - \mu \right) 
\left[ \left( 1-\mu^2 \right)
{\partial P_w(p,\mu,\mu^\prime) \over \partial \mu}
+\left( 1-{\mu^\prime}^2 \right) 
{\partial P_w(p,\mu,\mu^\prime) \over \partial \mu^\prime}
 \right]$$
$$ +{1 \over 8} p \left( V \over v \right)^2 \int_{-1}^1 {\rm d}\mu
\int_{-1}^1 {\rm d}\mu^\prime \left( {\mu^\prime}^2 - \mu^2 \right) 
{\partial P_w(p,\mu,\mu^\prime) \over \partial p}
 \qquad . \eqno(3.14)$$

\noindent
The analytic condition for vanishing of the term (3.14) is the symmetry
of the scattering probability $P_w$ with respect to the interchange of
$\mu$ and $\mu^\prime$

$$P_w(p,\mu,\mu^\prime) = P_w(p,\mu^\prime,\mu) \qquad . \eqno(3.15)$$

\noindent
If the principle of detailed balance is satisfied, the above condition
must also held (in general, the opposite is not always true). Then, for
two opposite scattering acts (${\bf p} \rightarrow {\bf p^\prime}$) and
(${\bf p^\prime} \rightarrow {\bf p}$) the respective momentum gain and
loss terms cancel each other exactly and any net energy change comes
from different particle densities near ${\bf p}$ and ${\bf p^\prime}$.
In most cases of ordinary scattering processes the condition (3.15) is
trivially satisfied and the momentum diffusion fully describes the
acceleration process (cf. Drury [3]). In order to illustrate the
physical reasons for non-vanishing regular term $\dot{p}_{reg}$, in the
following section a simple mechanical example with asymmetric particle
scattering is presented.

\section{An example with a non-vanishing regular
acceleration term}

As a simple illustration for the scattering process with the
cross-section showing the forward-backward asymmetry we consider the
elastically reflecting regular tetrahedrons (i.e. triangular pyramids
with all 4 walls being the equal triangles), moving always with one of
its bases directed forward and perpendicular to its direction of motion.
Let us measure all angles with respect to the base normal, pointing by
definition into $\mu = 1$. We consider particle reflections in the
tetrahedron rest frame. If we take a stream of particles hitting the
base with $\mu = -1$, then the reflected particles will be characterized
with pitch angle $\mu^\prime = 1$. However, if a stream hitting the
tetrahedron will have originally $\mu = 1$, then no one of particles
reflected from the inclined walls of tetrahedron will be scattered into
$\mu^\prime = -1$. This simple example demonstrates the breaking of the
condition (3.15). Next, let us derive the diffusion coefficient and,
both, the diffusive and the regular acceleration term existing for such
scattering centres. From numerical integration of Eq-s (3.12) and
(3.14), with the reflection probability defined by the tetrahedron
geometric shape, we obtain the following transport coefficients

$$\qquad D_p = 0.0813 \, \left( {V \over v}\right)^2 p^2  \hfill 
\eqno(4.1)$$
$$\qquad \dot{p}_{diff} = 0.1626 \, \left( {V \over v}\right)^2 p 
\hfill \eqno(4.2)$$
$$\qquad \dot{p}_{reg} = 0.0043 \, {V \over v}\, p 
+ 0.0004 \, \left({V\over v}\right)^2 \, p \qquad , \hfill \eqno(4.3)$$

\noindent
where we put $\Delta t = 1$.  In expressions (3.14) and (4.3) for
$\dot{p}_{reg}$ one may note the existence of terms representing both
the first-order and the second-order {\em regular} acceleration. The
occurrence of the term linear in $V/v$ is of particular interest in any
application of the theory to particle acceleration. In our example, for
small $V/v$, this term can dominate over another terms.

\section{Final remarks}

In the present work we considered conditions required for reduction of
the stochastic particle momentum scattering process into a pure
momentum diffusion. Our considerations were intended to translate this
rather general statement into the terms of scattering probability in
particle pitch angle and transport coefficients derived to the
second-order in $V/v$, as required by the Fokker-Planck approach. The
derived formula requires an equal probability in the wave frame to
scatter a particle from the pitch angle $\mu$ to $\mu^\prime$ and from
$ \mu^\prime$ to $\mu$. It is equivalent, after averaging over the
particle phase angle, to assumption of the detailed balance principle.
We also derived an expression for a regular acceleration term occurring
when the symmetry of the above probability is broken. In the previous
section we illustrate with a simple example the conditions allowing for
the regular acceleration term to exist.  The resulting coefficient
$\dot{p}_{reg}$ can consist of the linear in the scattering centre
velocity term as well as the second-order term proportional to
$(V/v)^2$. In astrophysical conditions one usually has $V \ll c$. Thus
the non-vanishing linear term could provide a fast way to accelerate
particles in the first-order process, with the regular term up to three
orders of magnitude larger than the diffusive term.

The importance of the present derivation depends on the possibility to
occur conditions allowing to break the scattering symmetry in the wave
frame (Ostrowski \& Siemieniec-Ozi\c{e}b{\l}o, in preparation). In this
respect one can consider regions of space with MHD turbulence
influenced by the strong external driving force  and leading to
high-amplitude waves. Thus, particles can be scattered by non-linear
asymmetric waves, allowing for different particle distributions
resulting from head-on and tail-on collisions. We expect such
conditions to occur near high Mach number shock fronts as well as in
boundary regions of high velocity jets penetrating through the cold
medium. 

{\bf Acknowledgements.}
The present work was initiated with a critical remark by Luke Drury.
Great thanks are to Peter Duffy for providing a part of his PhD thesis,
which proved to be very useful. The work was partly done during the
visit of MO to Max-Planck-Institut f\"ur Radioastronomie in Bonn. He is
grateful to Prof. Richard Wielebinski for hospitality. We acknowledge
critical comments of Reinhard Schlickeiser. The present work was
supported by the Polish Committee for Scientific Research through
grant PB 1117/P3/94/06.

}
\end{document}